\documentclass[]{aa} 

\usepackage{graphicx}
\usepackage{txfonts}
\def\lsim{\mathrel{\rlap{\lower4pt\hbox{\hskip1pt$\sim$}}
    \raise1pt\hbox{$<$}}}                
\def\gsim{\mathrel{\rlap{\lower4pt\hbox{\hskip1pt$\sim$}}
    \raise1pt\hbox{$>$}}}                
\begin{document}
%
  \title{Comparison of UV and high-energy ion irradiation of methanol:ammonia ice}

   \author{G. M. Mu\~noz Caro \inst{1}\fnmsep\thanks{send offprint requests to munozcg@cab.inta-csic.es}
          \thanks{Experiments performed at the Grand Acc\'el\'erateur National d'Ions Lourds (GANIL) in Caen, France. Part of the equipment used in this work has been financed by the French INSU-CNRS program
``Physique et Chimie du Milieu Interstellaire'' (PCMI).}
          \and
	E. Dartois
          \inst{2,3}          \and
          P. Boduch \inst{4}
          \and
          H. Rothard \inst{4}
          \and
          A. Domaracka  \inst{4}
          \and
          A. Jim\'enez-Escobar \inst{1}
          }
   \institute{
   Centro de Astrobiolog'a (CSIC-INTA), Carretera de Ajalvir, km 4, Torrej\'on de Ardoz, 28850 Madrid, Spain
   \and
CNRS-INSU, Institut d'Astrophysique Spatiale, UMR 8617, 91405 Orsay, France
\and
Universit\'e Paris Sud, Institut d'Astrophysique Spatiale, UMR 8617, b\^atiment 121, 91405 Orsay, France
\and
Centre de Recherche sur les Ions, les Mat\'eriaux et la Photonique (CEA/CNRS/ENSICAEN/Universit\'e de Caen-Basse Normandie),
CIMAP - CIRIL - Ganil, Boulevard Henri Becquerel, BP 5133, 14070 Caen Cedex 05, France
}

   \date{Received November 5, 2013; accepted April 30, 2014}

 
  \abstract
   {}
   {The main goal of this work is to compare the effects induced in ices of astrophysical relevance by high-energy ions, simulating cosmic rays, and by vacuum ultraviolet (UV) photons.}
   {This comparison relies on in situ infrared spectroscopy of 
   irradiated CH$_3$OH:NH$_3$ ice. Swift heavy ions were provided by the GANIL accelerator. The source of UV was a 
   microwave-stimulated hydrogen flow discharge lamp. The  deposited energy doses were similar for ion beams and UV photons to allow a direct comparison.}
   {A variety of organic species was detected during irradiation and later 
    during ice warm-up. These products are common to ion and UV 
    irradiation for doses up to a few tens of eV per molecule. Only the relative 
    abundance of the CO product, after ice irradiation, was clearly higher in the ion irradiation experiments.}
{For some ice mixture compositions, the irradiation products formed depend only weakly on the type of irradiation, swift heavy ions, or UV photons. This simplifies the chemical modeling of energetic ice processing in space.}
   \keywords{cosmic rays, dust, extinction -- 
                       ISM: molecules -- 
                       line: profiles, molecular processes
                       }
\titlerunning{UV vs. high energy ion irradiation of CH$_3$OH:NH$_3$ ice} 
   \maketitle
%

\section{Introduction}
\label{intro}
Dust grains covered by ice mantles in interstellar and circumstellar environments are exposed to photons and cosmic rays (ions). In dense interstellar cloud interiors, ice processing is mainly driven by cosmic rays and cosmic-ray-induced secondary UV photons (e.g., \cite{Cecchi1992,Shen2004}). The secondary UV spectrum is simulated well by the microwave-discharge hydrogen
flow lamp in the laboratory, with the exception of photon emission at wavelengths below 114 nm (because of the MgF$_2$ window cutoff, see  \cite{Cruz2014b} and references therein). As a result of this, a few molecules present in ice mantles such as CO are not directly dissociated in these experiments and other photon sources are required to explore the extreme-UV and soft X-ray processing of ices, as described for example by \cite{Wu2003} and \cite{Chen2013}.  
Ion irradiation of ices in the laboratory has long been dominated by moderate projectile energies (keV to MeV domain). The results obtained in these experiments were often extrapolated to higher energies (on the order of 1 GeV) for astrophysical ices. In recent years, some experiments have reached energies near 1 GeV
(\cite{Seperuelo2009}, 2010; \cite{Pilling2010a}; 2010b; \cite{Dartois2013}). 

Comparison between the irradiation products made by ions and UV photons is only possible for a reduced number of ice mixtures. 
A recent introduction to this topic is provided by \cite{Islam2014}. Below, we only refer to a few publications in relation to our work. 
The products of 0.8 MeV protons and UV irradiation were compared for H$_2$O ice that contained CO$_2$, 
CH$_3$OH, CH$_4$, HCN, or NH$_3$ by \cite{Moore2001}, \cite{Gerakines2001}, and \cite{Gerakines2004}. These authors observed the formation of similar products regardless of the nature of the radiation (photons or ions). Recently, CH$_3$OH:N$_2$ ice was found to lead to N-bearing products after 200 keV H$^+$ ion bombardment, while UV irradiation did not produce such species according to \cite{Islam2014}, because of the low photon absorption of N$_2$ in the vacuum-UV range (\cite{Cruz2014a}).
UV irradiation experiments commonly used H$_2$O-rich ice mixtures that contained NH$_3$ and often CH$_3$OH 
(e.g., \cite{Agarwal1985}; \cite{Briggs1992}; \cite{Bernstein1995}; \cite{Bernstein2002}; \cite{MC2002}; \cite{MCS2003}; \cite{Meierhenrich2005}; \cite{Nuevo2006}).
The infrared spectrum of a room-temperature residue made from ion irradiation of an interstellar ice analog, dominated by H$_2$O and containing CH$_3$OH and NH$_3$ was to our knowledge not reported in the literature. In addition to the references cited above, a residue spectrum that resulted from 0.8 MeV proton bombardment, provided by Marla Moore, is reported in \cite{Pend&Alla2002},
but it was made starting from a CH$_4$:NH$_3$ ice composition. Similarly, 
Strazzulla et al. (2001) 
showed the infrared spectrum of the room-temperature residue obtained from ion irradiation of H$_2$O:CH$_4$:N$_2$ ice. The room-temperature residue spectrum of 
H$_2$O:CH$_3$OH = 1:0.6 ice made from 1 MeV proton irradiation was reported by Moore et al. (1996). 
This spectrum looks similar to that obtained from 
UV irradiation of a similar ice, H$_2$O:CH$_3$OH = 1:1, shown in 
Fig.~2 (bottom panel) of \cite{Mun&Dar2009}, although there are some 
differences in the fingerprint region. The main difference is the apparent absence or small 
depth of the absorption around 1029 cm$^{-1}$ in the spectrum of Moore et al. (1996). 
It would therefore be interesting to compare the organic products obtained from 
ion irradiation of H$_2$O:CH$_3$OH:NH$_3$:CO:CO$_2$ ices with those made by UV 
irradiation that were previously published. The deposited energy 
(eV cm$^{-2}$) in both types of experiments should also be similar for a proper 
comparison. 

The goal of this work is twofold. First, we aim to reproduce the irradiation of ice mantles to study the formation of complex organic molecules by high-energy cosmic rays in dense clouds and circumstellar regions, using the most energetic ions available experimentally. Second, we compare this experiment with a UV-irradiation experiment that uses the same ice mixture and deposited energy dose. The 
CH$_3$OH:NH$_3$ ice mixture was selected because several organic products were identified in UV-irradiation experiments, 
see \cite{MCS2003}, and the formation of products can be 
elucidated using a relatively small set of reactions. Addition of H$_2$O to this ice mixture leads to a more complex chemistry (\cite{MCS2003} and \cite{Oberg2010}) and is left for future work.    

\begin{table*}[htdp]
\begin{center}
\caption{New bands attributed to irradiation products}
\begin{center}
\begin{tabular}{l l l l l}
\hline
position$\rm^{a}$ ($\rm cm^{-1}$)		& Assignment			& vibration mode	&UV after dep.	& Zn (620 MeV)	\\
\hline
2340 & CO$_2$ 			   & CO str. & $\times$ &$\times$\\
2160 & OCN$^-$ 	           & CN str. & $\times$ &$\times$\\
2138 & CO 			   & CO str. & $\times$ &$\times$\\
1740 & C=O ester/aldehyde 	   & CO str. & $\times$ &$\times$\\
1720 & H$_2$CO 		   & CO str. & $\times$ &$\times$\\
1694 & HCONH$_2$ ? 		   & CO str. & $\times$ &$\times$\\
1587 & COO$^-$ in carb. ac. salts$\rm^{b,c}$  & COO$^-$ asym. str. & $\times$ & $\times$\\
1498 & H$_2$CO 		   & CH$_2$ scis. & $\times$ &$\times$\\
1385 & CH$_3$ groups                      & CH$_3$ sym. def. & $\times$ &$\times$\\
1347 & COO$^-$ in carb. ac. salts$\rm^{b,c}$  & COO$^-$ sym. str. & $\times$ &$\times$\\
1303 & CH$_4$ 		   & def. & $\times$ &$\times$\\
\hline
\end{tabular}
\end{center}
$\rm^{a}$ Position varies sligthly because of the interaction of species within the matrix; $\rm^{b}$  Mu\~noz Caro \& Schutte (2003) ; $\rm^{c}$ Nuevo et al. (2006).
\label{positions}
\end{center}
\end{table*}
%




\section{Experiments}

\subsection*{Heavy ions}
Ion-irradiation experiments were performed at the heavy-ion accelerator Grand Acc\'el\'erateur National d'Ions Lourds (GANIL).
Zn$^{26+}$ projectiles were accelerated to 620 MeV on the medium energy exit beam line in November 2010 and Ne$^{6+}$ projectiles were accelerated to 19.6 MeV  in June 2012 at the IRRSUD beam line, respectively.
A high-vacuum chamber, with base pressure in the 10$^{-7}$ mbar range at room temperature, holding an infrared-transmitting substrate cryocooled at 10 K, on top of which the ice mixture 
was condensed, was mounted on these beam lines. The ice films were produced by placing the cold window substrate in front of a deposition line where gas mixtures are prepared in the appropriate proportions. The film thickness was chosen to give a high band contrast  with respect to the infrared absorptions, without saturating the bands. At such thicknesses, the ion beam passes through the film with an almost constant energy deposition of about 
250 eV/$\AA$ (Zn projectile) and 
130 eV/$\AA$ (Ne projectile), for a mean density of 0.73 g cm$^{-3}$ for the NH$_3$:CH$_3$OH ice mixture (assuming a density of 0.67~g cm$^{-3}$ for NH$_3$ ice (\cite{Satorre2013}) and the liquid value of 0.79~g/cm$^{-3}$ for CH$_3$OH).
A Nicolet FTIR spectrometer (Magna 550) with a spectral resolution of 
1 cm$^{-1}$ was used. The evolution of the spectra was recorded at several 
fluences.  For more experimental details, see Seperuelo Duarte et al. (2009).

\subsection*{UV photons}

The UV-irradiation experiments were perfomed using the InterStellar 
Astrochemistry Chamber (ISAC), an ultra-high vacuum setup 
($2.5 - 4.0 \times 10^{-11}$ mbar) where an ice sample is made by deposition 
of a gas mixture onto the cold finger of a closed-cycle helium cryostat at 8 K.
The design of the gas line incorporates electrovalves controlled by a 
quadrupole mass spectrometer, QMS (Pfeiffer Vacuum, Prisma QMS 200). It 
allows controling the relative proportions of the gas mixture components 
before and during the deposition.      
Samples were UV-irradiated using a microwave-stimulated hydrogen flow 
discharge lamp that provides a flux of 
2.5 $\times$ 10$^{14}$ photons cm$^{-2}$ s$^{-1}$ at 
the sample position with an average photon energy of 8.8 eV. After 
irradiation, the ice is warmed using a constant heating ramp of 1--2 K 
min$^{-1}$. The ice is monitored by in situ transmittance Fourier transform 
infrared (FTIR) spectroscopy (Bruker VERTEX 70), while the volatile species 
are monitored by a second QMS equipped with a Channeltron detector 
(Pfeiffer Vacuum, Prisma QMS 200). 
The evolution of the ice was monitored by FTIR spectroscopy 
in transmittance at a spectral resolution of 2 cm$^{-1}$, taking spectra 
before and after the irradiation. For a more detailed description of ISAC, we 
refer to \cite{MC2010}.

\begin{figure}[htbp]
\begin{center}
\includegraphics[width=\columnwidth]{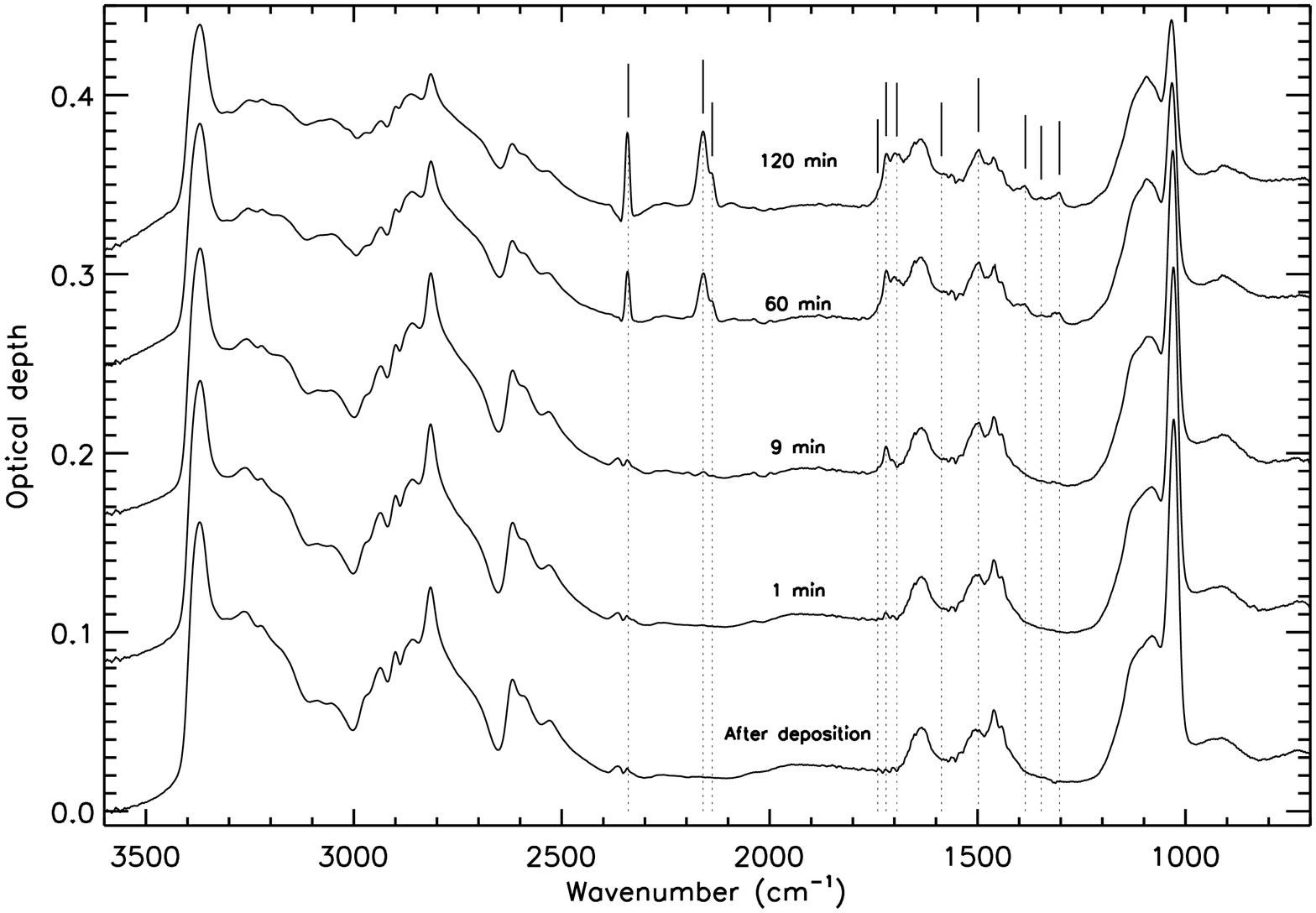}
\includegraphics[width=\columnwidth]{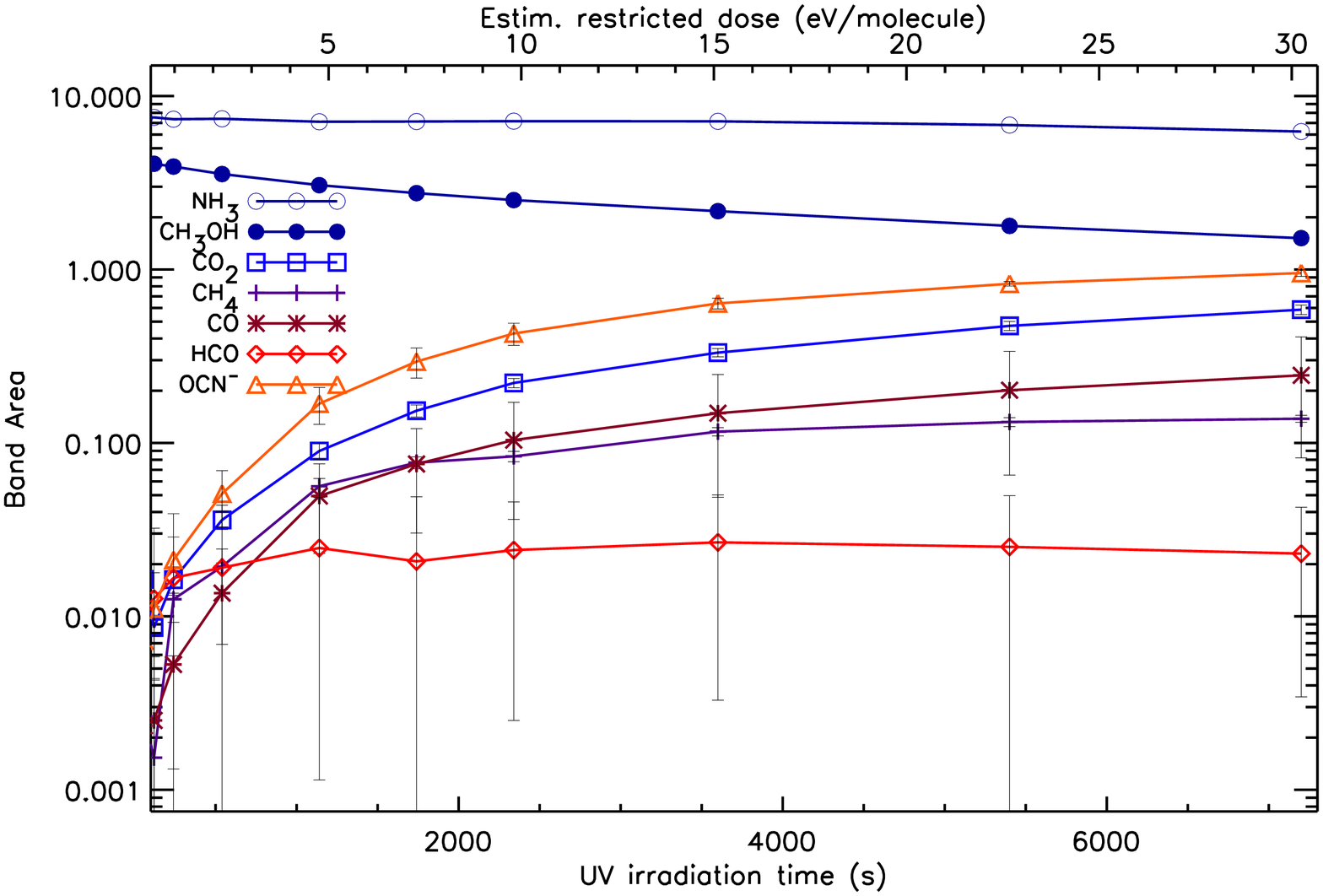}
\caption{UV irradiation of CH$_3$OH:NH$_3$ ice mixture. The top panel shows the IR spectra at various irradiation times after deposition. The bottom panel shows the decrease of the starting ice bands, corresponding to NH$_3$ and CH$_3$OH, and the growth of various bands attributed to irradiation products: CO$_2$, CH$_4$, CO, HCO, and OCN$^-$.} 
\label{UVfig1}
\end{center}
\end{figure}
\begin{figure}[htbp]
\begin{center}
\includegraphics[width=\columnwidth]{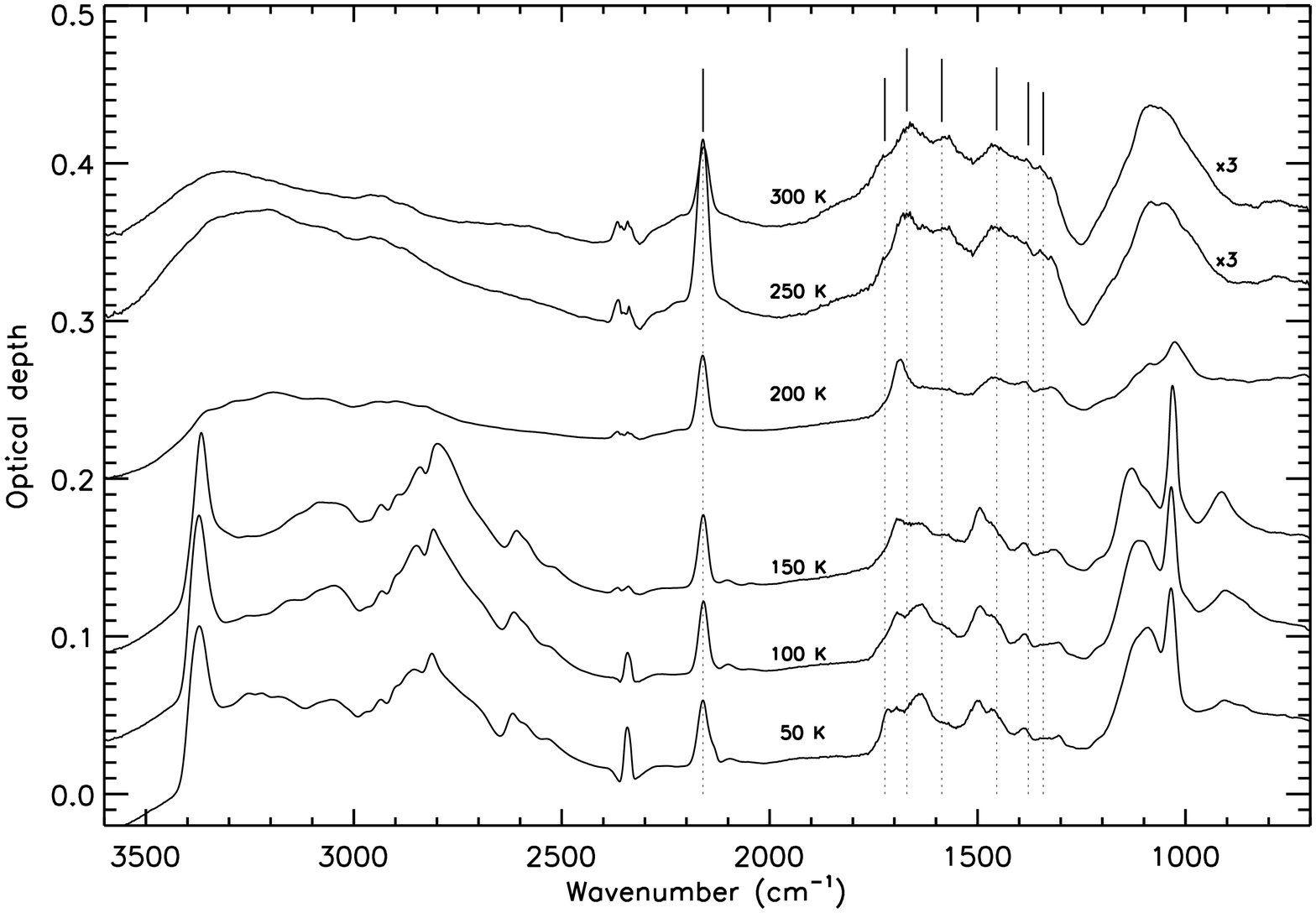}
\caption{Warm-up spectra of UV-irradiated CH$_3$OH:NH$_3$ ice mixture. 
The vertical line positions from 2160 cm$^{-1}$ to 1342 cm$^{-1}$ are presented and assigned to different vibrations in 
Table \ref{residue_carriers}. They correspond to the bands identified in the ion-irradiated mixture residues observed at 300 K
for direct comparison with the UV-produced residue at the same temperature.}
\label{UVresidue}
\end{center}
\end{figure}

%

\begin{figure}[htbp]
\begin{center}
\includegraphics[width=\columnwidth]{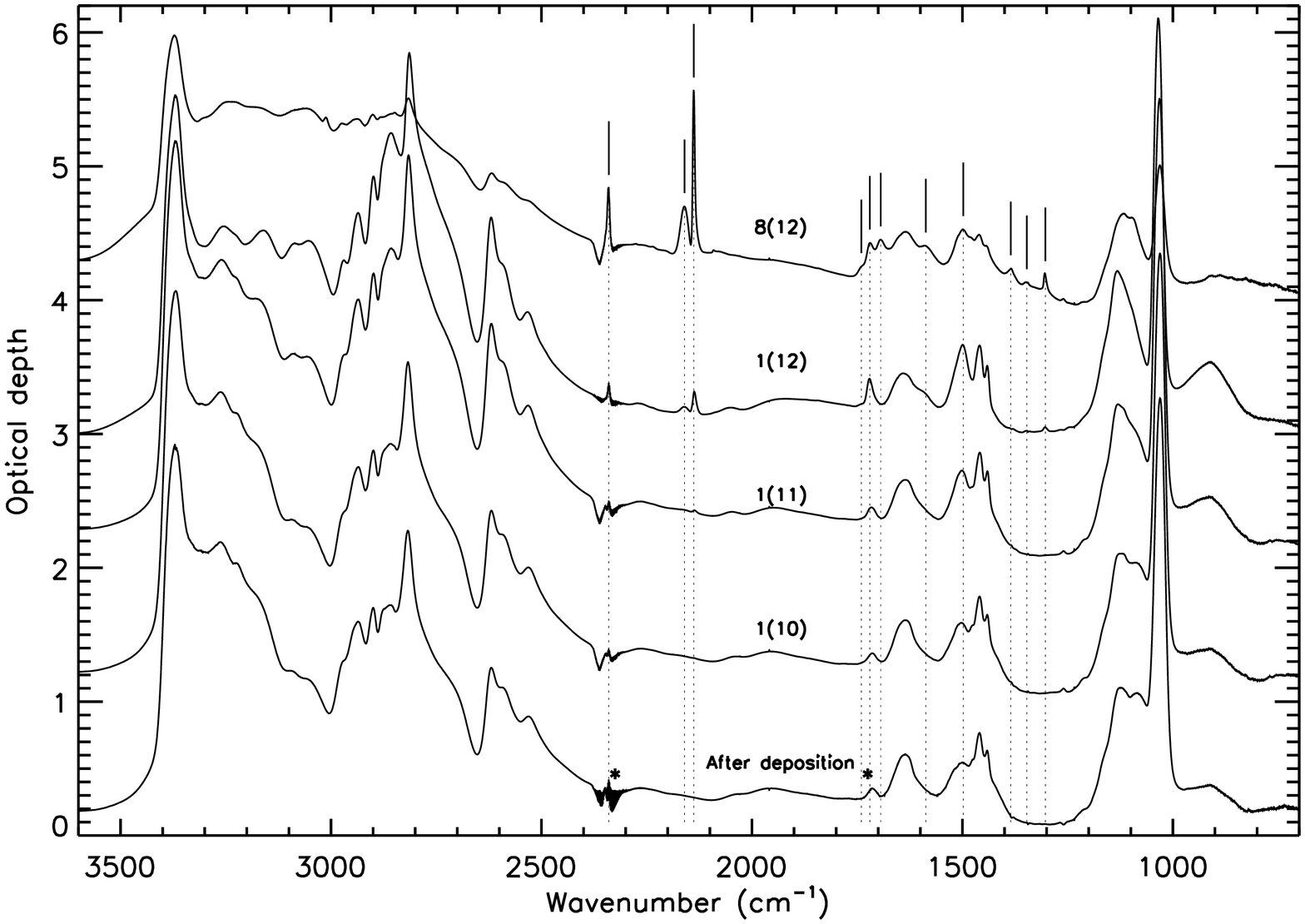}
\includegraphics[width=\columnwidth]{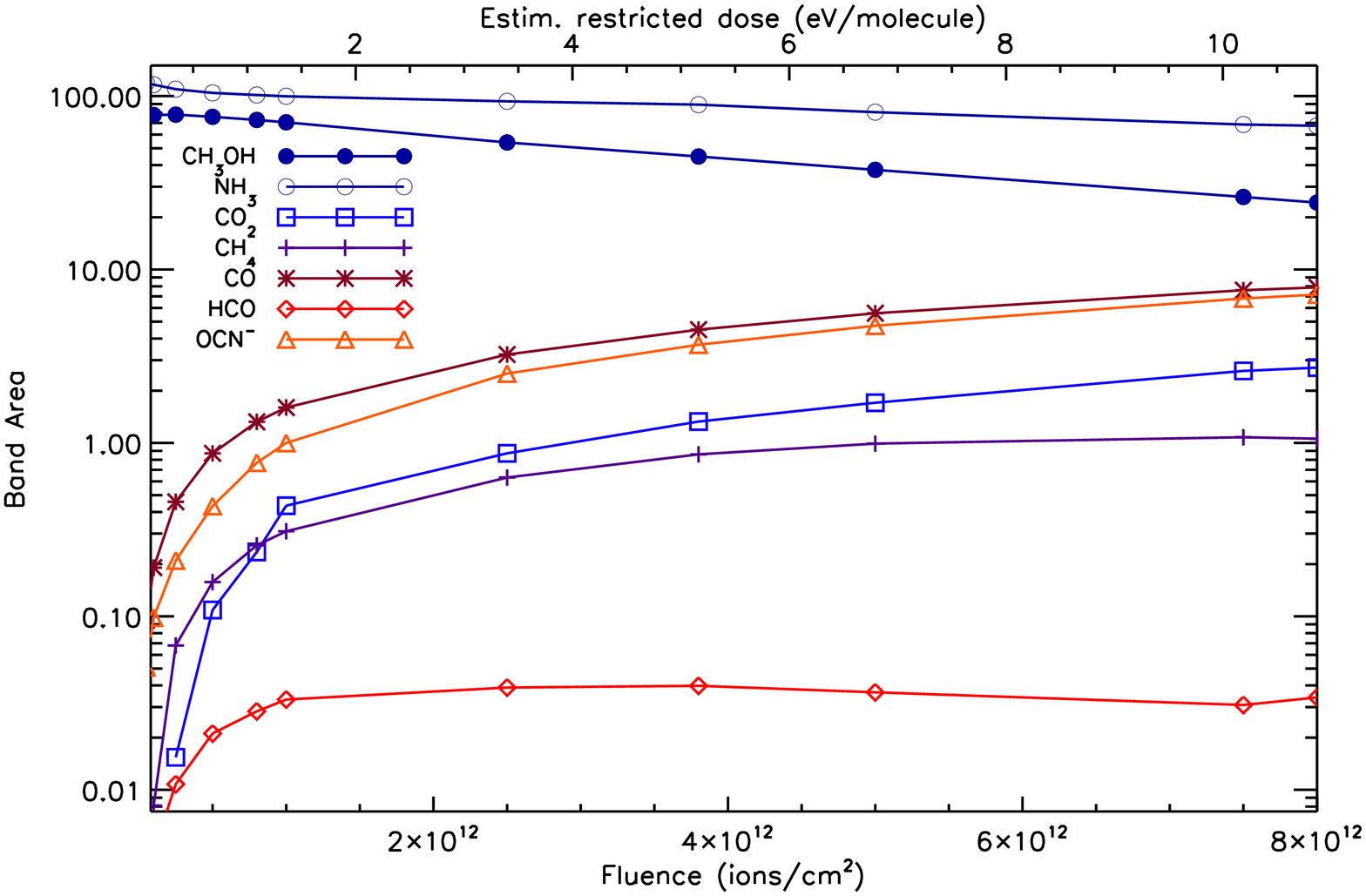}
\caption{620 MeV Zn ion irradiation of CH$_3$OH:NH$_3$ ice mixture.
The top panel shows the IR spectra at various irradiation intervals after deposition. The bottom panel shows the decrease of the starting ice bands, corresponding to NH$_3$ and CH$_3$OH, and the growth of various bands attributed to irradiation products: CO$_2$, CH$_4$, CO, HCO, and OCN$^-$.}
\label{ion_fig3}
\end{center}
\end{figure}
\begin{figure}[htbp]
\begin{center}
\includegraphics[width=\columnwidth]{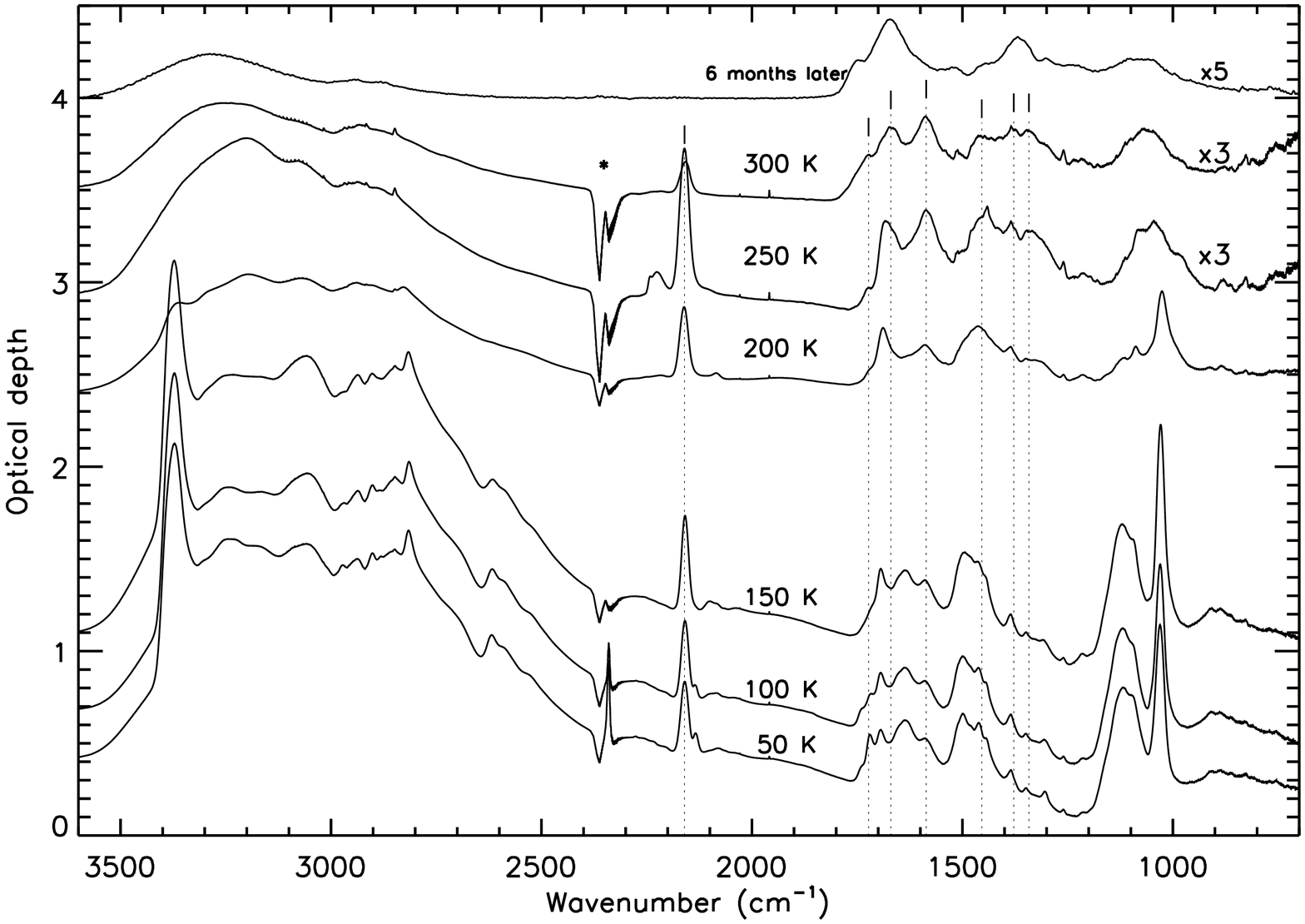}
\caption{Warm-up spectra of 620 MeV Zn ion irradiated CH$_3$OH:NH$_3$ ice mixture. The vertical line positions from 
2160 cm$^{-1}$ to 1342 cm$^{-1}$ are presented and assigned to different vibrations in Table \ref{residue_carriers}. }
\label{ion_residue}
\end{center}
\end{figure}

\begin{figure}[htbp]
\begin{center}
\includegraphics[width=\columnwidth]{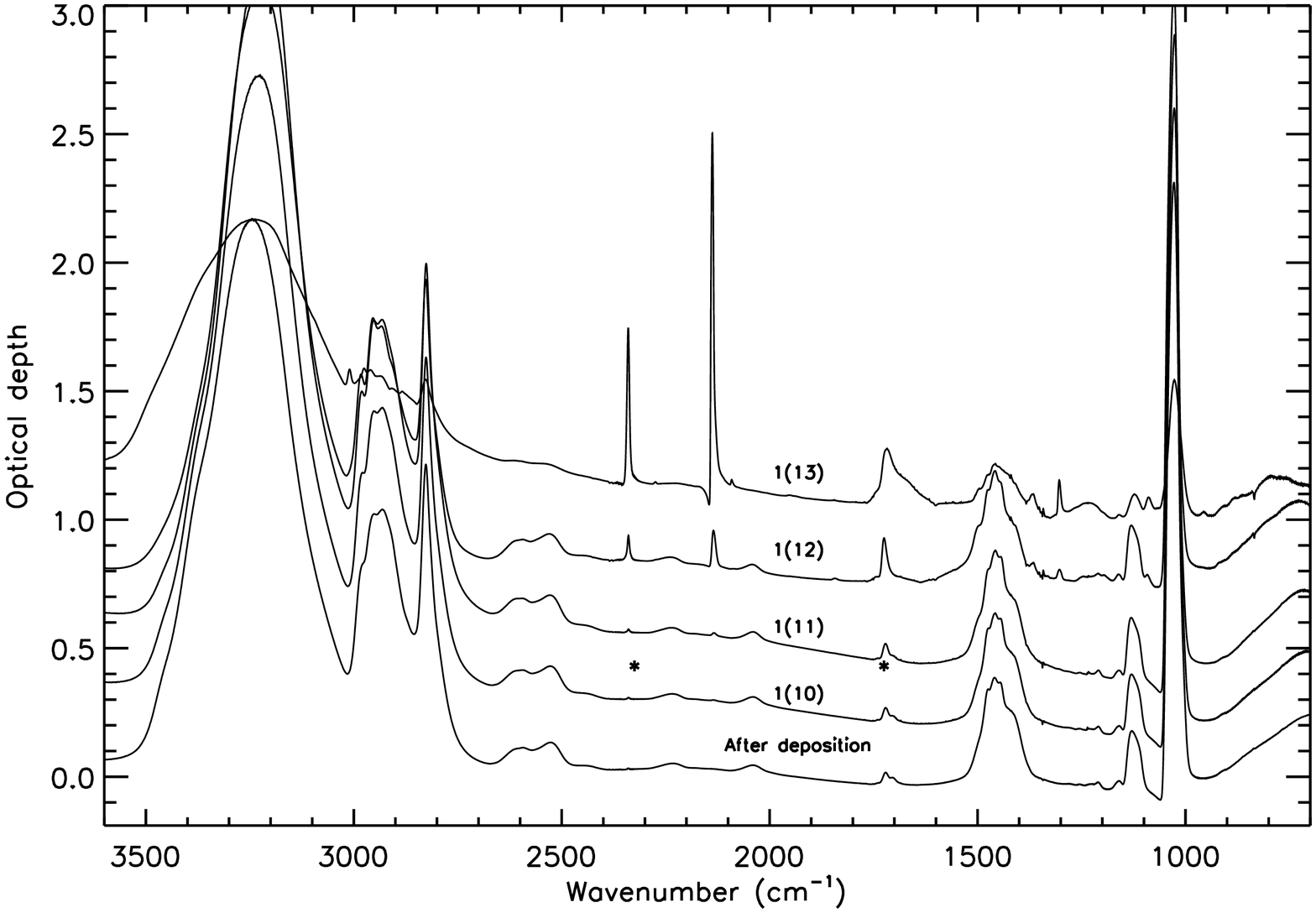}
\includegraphics[width=\columnwidth]{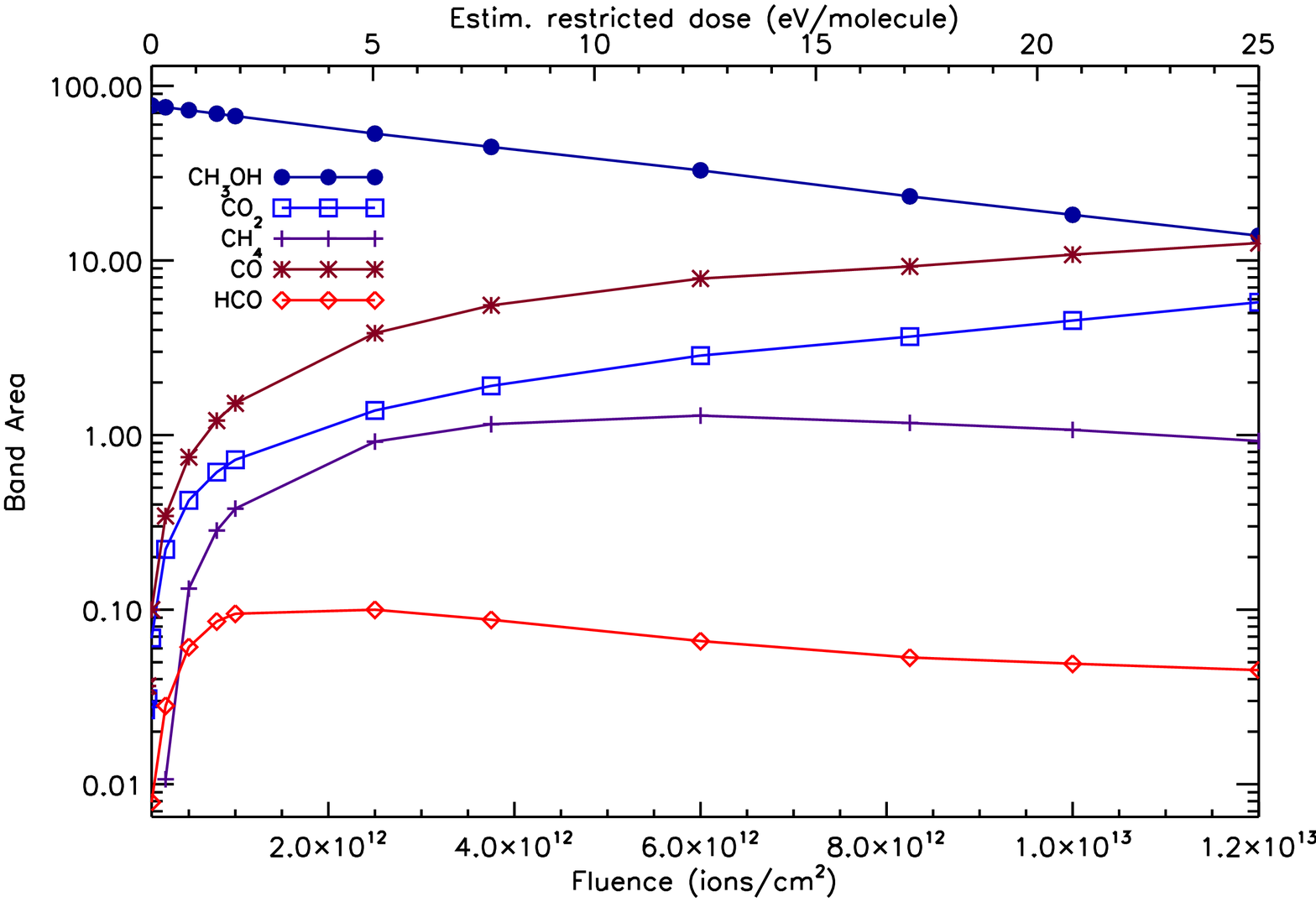}
\caption{620 MeV Zn ion irradiation of pure CH$_3$OH ice.
The top panel shows the IR spectra at various irradiation intervals after deposition. The bottom panel shows the decrease of the starting ice CH$_3$OH band and the growth of various bands attributed to irradiation products: CO$_2$, CH$_4$, CO, and HCO.}
\label{fig_ch3oh_zn620_1}
\end{center}
\end{figure}
\begin{figure}[htbp]
\begin{center}
\includegraphics[width=\columnwidth]{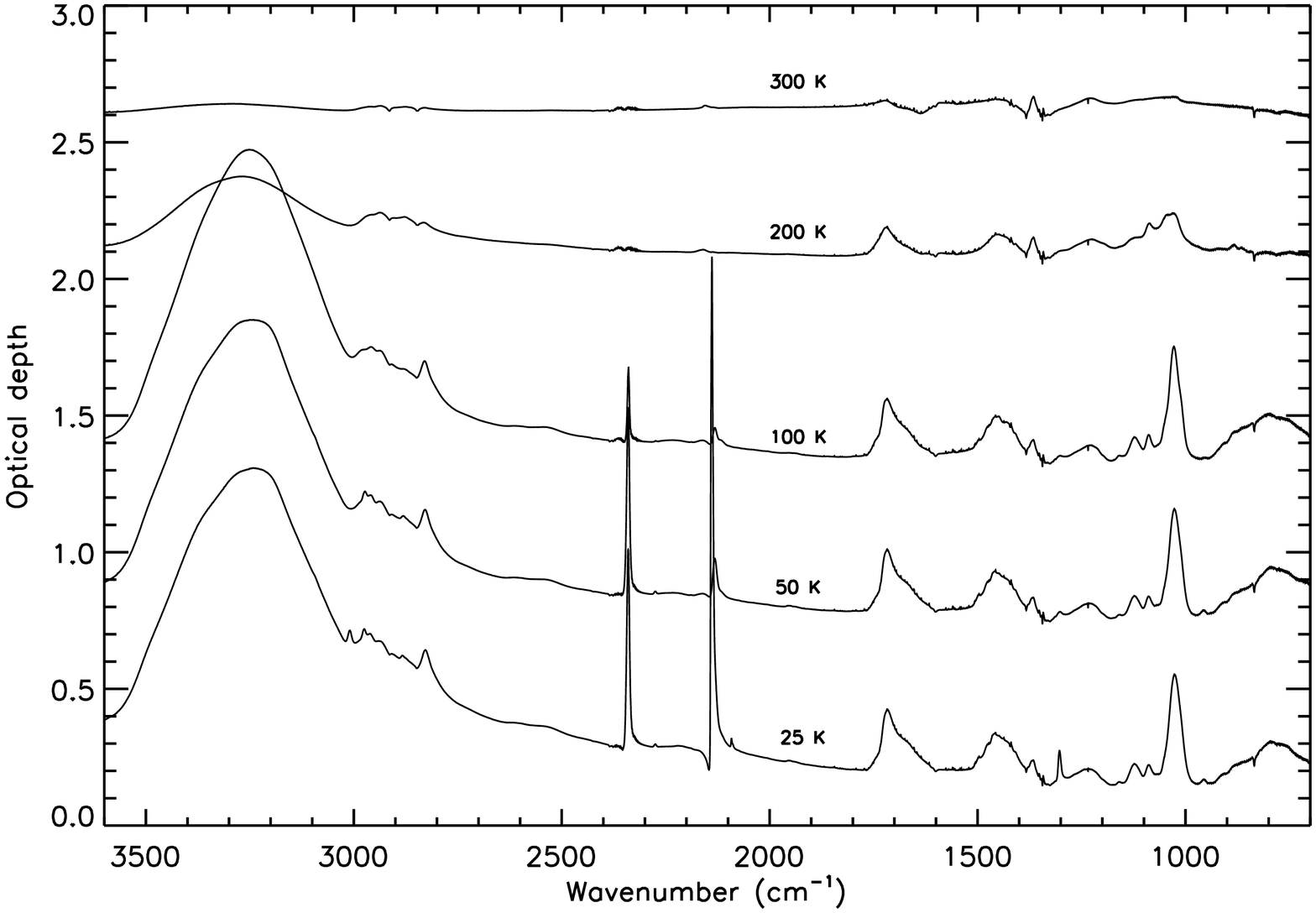}
\caption{Warm-up of 620 MeV Zn ion irradiated pure CH$_3$OH ice.}
\label{fig_ch3oh_zn620_2}
\end{center}
\end{figure}

\begin{figure}[htbp]
\begin{center}
\includegraphics[width=\columnwidth]{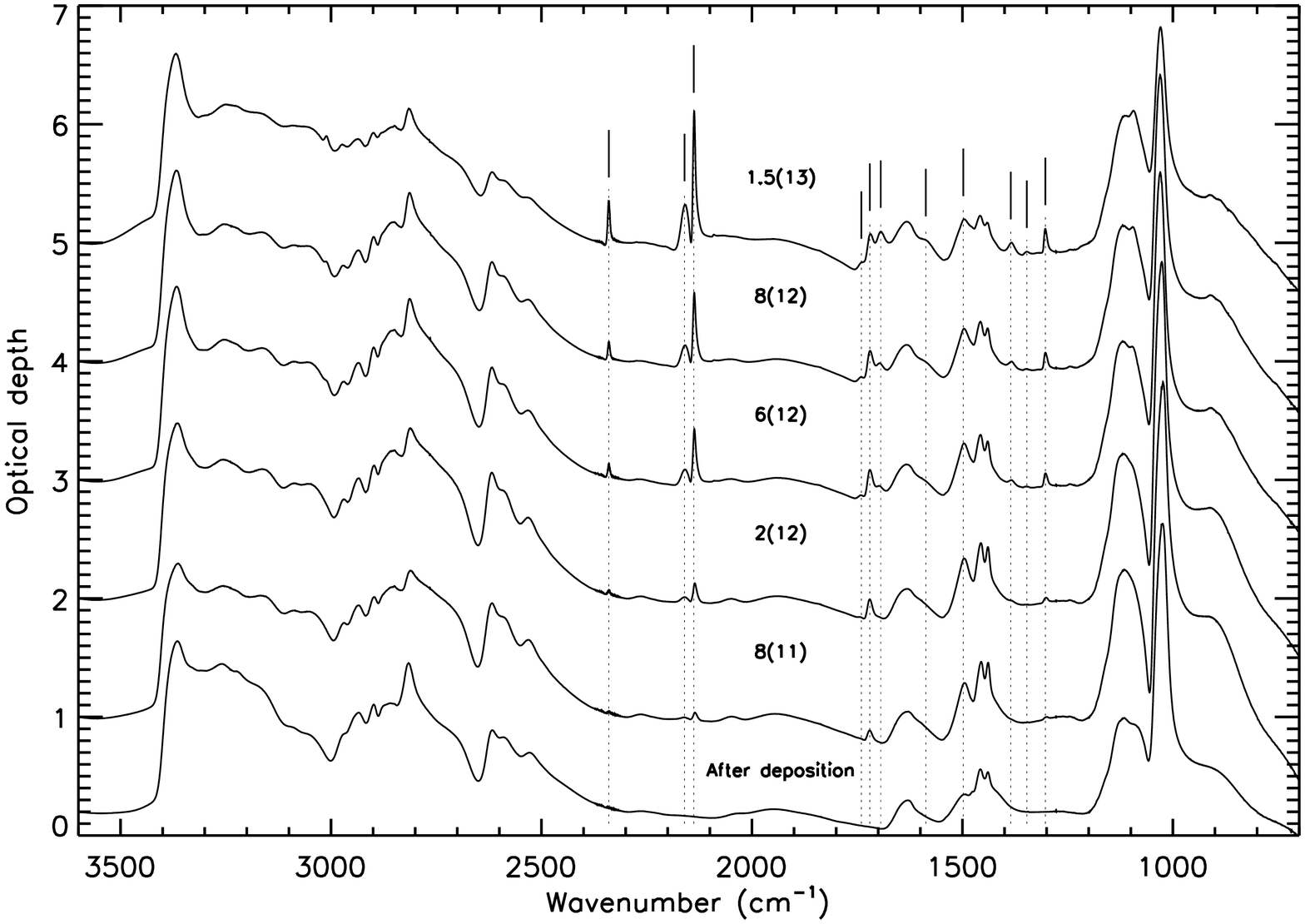}
\includegraphics[width=\columnwidth]{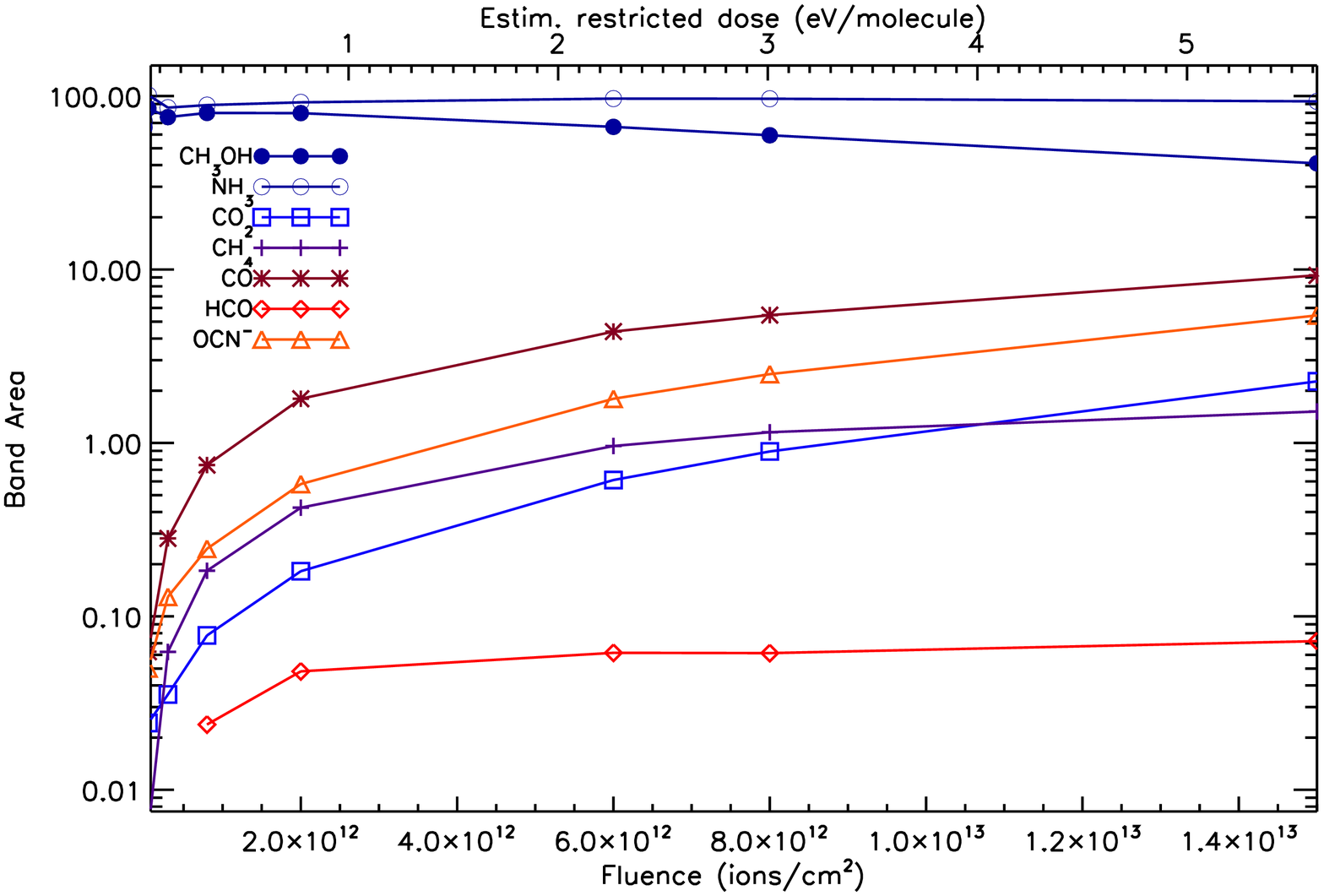}
\caption{19.6 MeV Ne ion irradiation of CH$_3$OH:NH$_3$ ice mixture.
The top panel shows the IR spectra at various irradiation intervals after deposition. The bottom panel shows the decrease of the starting ice bands, corresponding to CH$_3$OH and NH$_3$, and the growth of various bands attributed to irradiation products: CO$_2$, CH$_4$, CO, HCO, and OCN$^-$.
}
\label{Fig1_Ne}
\end{center}
\end{figure}
\begin{figure}[htbp]
\begin{center}
\includegraphics[width=\columnwidth]{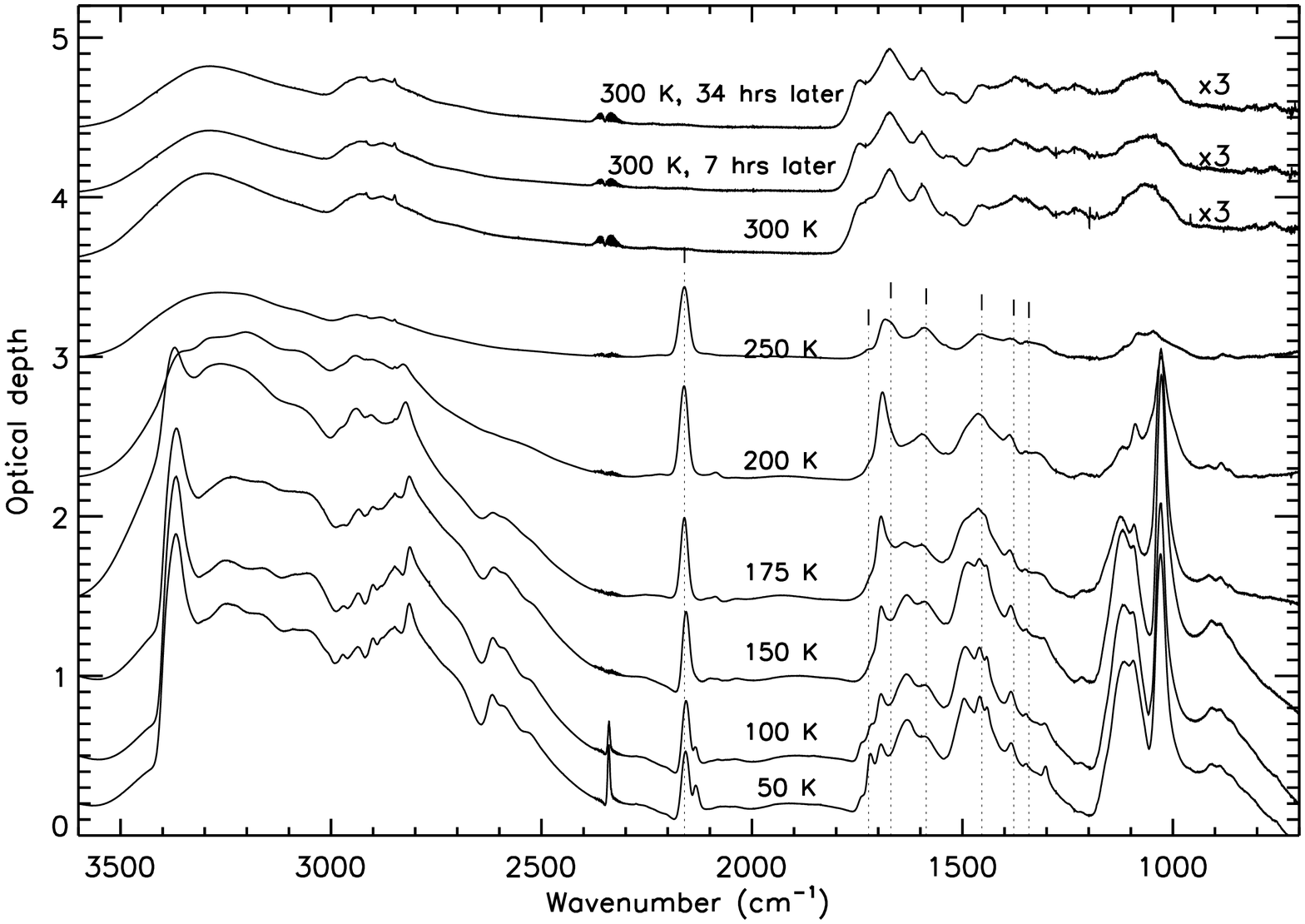}
\caption{Warm-up spectra of 19.6 MeV Ne ion irradiated CH$_3$OH:NH$_3$ ice mixture.
The vertical line positions from 2160 cm$^{-1}$ to 1342 cm$^{-1}$ are presented and assigned to different vibrations in 
Table \ref{residue_carriers}. 
}
\label{Fig2_Ne}
\end{center}
\end{figure}

\section{Results}
\label{results}
To compare the deposited energies involved across the different 
experiments, the ion fluence was converted into restricted dose (eV/molecule) using the Stopping and Range of Ions in Matter (SRIM, \cite{Ziegler2008})
calculations  for a NH$_3$:CH$_3$OH = 1:1 ice mixture with an overall density 
of 0.73 g cm$^{-3}$.

The total column density of the ice deposited in the UV irradiation experiment, N(NH$_3$ + CH$_3$OH) = 6 $\times$ 10$^{17}$ molecules cm$^{-2}$, was estimated from integration of the CO stretching (${\nu}_8$) band of methanol absorbing at 1026 
cm$^{-1}$ (\cite{d'Hend&Allam1986}), 1.8 $\times$ 10$^{-17}$ cm molecule$^{-1}$. 
The average UV absorption cross-section of the ice mixture is about 7 $\times$ 10$^{-18}$ cm$^{2}$, based on the values for ammonia ice, $\sim$ 5 $\times$ 10$^{-18}$ cm$^{2}$ (\cite{Mason2006}, and \cite{Cruz2014b}), and methanol ice, $\sim$ 7 $\times$ 10$^{-18}$ cm$^{2}$ (\cite{Cruz2014b}). Therefore, about 97\% of 
the UV photons are absorbed in the ice. For a UV lamp delivering a flux of 2.5 $\times$ 10$^{14}$ photons cm$^{-2}$ s$^{-1}$, the average deposited energy is obtained by dividing the total number of absorbed photons by the column density of the deposited ice mixture; it corresponds to $\sim$ 0.21 eV per molecule per minute of irradiation.
%

The top panel of Fig.~\ref{UVfig1} displays the infrared spectra of a 
NH$_3$:CH$_3$OH = 1:1 ice mixture after deposition and during UV irradiation.
The total UV irradiation time is given in minutes. A decrease of the initial ice
absorptions during UV irradiation is accompanied by the emergence of new 
absorptions corresponding to the products presented in Table~\ref{positions}.
The fourth column indicates the various products in the UV 
experiment. The CO, CO$_2$, H$_2$CO, and CH$_4$ products are also formed 
by UV irradiation of pure CH$_3$OH ice (\cite{Gerakines1996}).
The evolution of the integrated absorption bands for the main 
UV-irradiation products is shown in the bottom panel of Fig.~\ref{UVfig1}.    
Fig.~\ref{UVresidue} displays the warm-up spectra of the irradiated ice.
The room temperature spectrum at 300 K is associated with an organic refractory
residue with notable absorption bands at 
about 3600-2500 (broad), 2929, 2876, 2160, 1723, 1670, 1586, 1454, 1378, 1342, and 1050-1060(broad)~cm$ ^{-1}$. The molecular carriers of these bands are given in Table \ref{residue_carriers}. 
\begin{center}
\begin{table*}      
\caption{Assigned feature carriers of the IR-residue spectrum formed by UV irradiation of NH$_3$:CH$_3$OH = 1:1 ice.}
\label{residue_carriers}
\begin{tabular}{cccccc} \hline

Position  & Assignment & Vibration mode & UV after dep.	& Zn (620 MeV)\\ 
cm$^{-1}$  &            &                &               &             \\  \hline  
3600-2300 & R-COOH, alcohols, NH$_4^+$$^a$ & OH str., NH str. & $\times$ & $\times$ \\
2930      & -CH$_2$OH$^b$ & 2${\nu}_{19}$ antisym. -CH$_2$ str. & $\times$ & $\times$\\
2875      & -CH$_2$OH$^b$, NH$_{4}^{+}$$^a$ & ${\nu}_{18}$ sym. CH$_2$ str., 2${\nu}_4$ of NH$_{4}^{+}$$^a$ & $\times$ & $\times$ \\
2160      & OCN$^-$  & CN str.  & $\times$ & $\times$ \\
1723      & Aldehydes & C=O str. & $\times$ & $\times$ \\
1670      & Amides & C=O str. & $\times$ & $\times$ \\ 
1586      & COO$^-$ in carboxylic acid salts $^{c,d}$ & COO$^-$ antisym. str. & $\times$ & $\times$ \\
1454      & NH$_{4}^{+}$$^a$, TMTH$^+$$^e$ & ${\nu}_4$$^a$ & $\times$ & $\times$ \\ 
1378      & CH$_3$ groups & CH scissoring $^a$ & $\times$ & $\times$ \\
1342      & COO$^-$ in carboxylic acid salts $^{c,d}$ & COO$^-$ sym. str.             & $\times$ & $\times$ \\
1098      & POM-like species  $^{c}$  &                                   & $\times$ & $\times$ \\ 
1050      & CH2-OH in primary alcohols    & C-O str. & $\times$ & $\times$ \\  
%
%
\hline
\end{tabular}
\begin{list}{}{}
\item[$^a$] Wagner \& Hornig (1950)
\item[$^b$] Mu\~noz Caro \& Dartois (2009) 
\item[$^c$] Mu\~noz Caro \& Schutte (2003) 
\item[$^d$] Nuevo et al. (2006) 
\item[$^e$] Vinogradoff et al. (2013), and references therein. TMTH$^+$ is the protonated
form of the trimethylenetriamine, C$_3$H$_{10}$N$_3$ 
\end{list}
\end{table*}
\end{center}

The top panel of Fig.~\ref{ion_fig3} displays the infrared spectra of a 
NH$_3$:CH$_3$OH = 1:1 ice mixture after deposition and during irradiation
using 620 MeV Zn ions.
The corresponding fluence, defined as the number of incident ions per cm$^{2}$, is given for each spectrum. A decrease of the initial ice
absorptions during irradiation is accompanied by the emergence of new 
absorptions corresponding to the products presented in Table~\ref{positions}.
The fifth column of Table~\ref{positions} indicates the various products formed during irradiation. 
The CO, CO$_2$, H$_2$CO, and CH$_4$ products are also formed by fast heavy ion-irradiation of pure CH$_3$OH ice (\cite{deBarros2011}). 
A band near 2234 cm$ ^{-1}$ corresponding to N$_2$O was not observed in our experiment. According to Pilling et al. (2010a), 
N$_2$ molecules are not easily formed from NH$_3$ and they are required as an intermediate step to synthesize N$_2$O.
The evolution of the integrated absorption bands for the main 
irradiation products is shown in the bottom panel of Fig.~\ref{ion_fig3}.    
Fig.~\ref{ion_residue} displays the warm-up spectra of the ion irradiated ice.
The room-temperature spectrum at 300 K is associated with an organic refractory
residue with absorption bands at 
3600-2500 (broad), 2929, 2876, 2160, 1723, 1670, 1586, 1454, 1378, 1342, and 1050-1060 (broad)
cm$ ^{-1}$. Vertical dotted lines indicate these positions in the figure.
The molecular carriers of these bands are given in Table \ref{residue_carriers}. 

A pure CH$_3$OH sample was irradiated with the same ions, explaining the formation of some of the species observed in the irradiated mixture (Fig.~\ref{fig_ch3oh_zn620_1}), but not leading to a significant residue during warm-up 
(Fig.~\ref{fig_ch3oh_zn620_2}).

To complete the set, an irradiation experiment was conducted with 19.6 MeV Ne ions on a NH$_3$:CH$_3$OH = 1:1 ice mixture, to repeat the first experiment using a slightly different stopping power. The infrared spectra during irradiation and the evolution of species with the restricted dose are displayed in  Fig.~\ref{Fig1_Ne}. A similar residue (Fig.~\ref{Fig2_Ne}) is obtained during warm-up.

\section{Discussion and astrophysical implications}
\label{discussion}

The residue spectrum obtained from heavy-ion irradiation 
(46 MeV $^{58}$Ni$^{13+}$) of H$_2$O:NH$_3$:CO = 1:0.6:0.4 (\cite{Pilling2010a})
differs from the organic refractory residue produced by UV irradiation of an ice with a similar 
composition reported in \cite{MCS2003}. In addition to the different type of irradiation, ions or UV, 
these differences might be due to other factors. First, the experimental protocol of these 
experiments was different: the ion experiment involved irradiation after 
deposition of the ice, while the UV experiment was performed by simultaneous 
deposition and irradiation. We found that a new residue made in ISAC by 
simultaneous deposition and UV irradiation of NH$_3$:CH$_3$OH = 1:1 ice, 
not shown, was very similar to that reported in Fig.~8 of \cite{MCS2003} and clearly differs from the one reported 
in Fig.~\ref{UVresidue} of our present work, made by irradiation after the ice deposition was complete, thus avoiding 
irradiation of the molecules in the gas phase before they accrete onto the cold finger. Second, the deposited energy dose in this ion experiment, 5 $\times$ 10$^{11}$ (46 MeV Ni ions) cm$^{-2}$, 
was also different from the UV experiment, 0.24 photon molecule$^{-1}$. Third, the dissociation 
of the CO molecule by ions, which is not possible for UV photons of energies
$\le$ 10.2 eV in a direct way, might also lead to different products. However, this 
needs to be tested experimentally. Breaking of the CO occurs if a UV-excited CO 
molecule can react with another CO to form CO$_2$ + C, a process with a very 
low efficiency for pure CO ice; see, e.g., Mu\~{n}oz Caro et al. (2010).      

The evolution of the abundances for the simple products detected during ion irradiation is similar regardless of the ion source, 620 MeV Zn or 19.6 MeV Ne ions, see Figs.~\ref{ion_fig3} and \ref{Fig1_Ne}. Similar products and reaction kinetics during ion exposure were also observed in the  CH$_3$OH:NH$_3$ and pure CH$_3$OH experiments using 620 MeV Zn ions, with the obvious exception of the OCN$^-$ species, Figs.~\ref{ion_fig3} and \ref{fig_ch3oh_zn620_1}. The relative abundance of CO in the irradiated ice was higher in the ion experiments,  Figs.~\ref{ion_fig3} and \ref{Fig1_Ne}, than in the UV experiment, 
Fig.~\ref{UVfig1}, but for comparable restricted dose values the other products displayed similar trends and relative abundances.

In our experiments with CH$_3$OH:NH$_3$ ice, UV, and swift-ion irradiation give, to first order, rise to 
the same general trend with respect to the residues formed after
warm-up to room temperature of the irradiated ices, when normalized to the same 
dose per molecule. Solid CH$_3$OH and NH$_3$ are easily dissociated by vacuum-UV photons. 
At low temperature, the higher energy density deposited by ions produces small molecules with a higher efficiency than UV photons for species with a higher dissociation energy threshold, or because a cascade of ionizations occur in a small volume simultaneously, see, for example, Table 3 of Seperuelo Duarte et al. (2010) for CO formation cross-sections. On the other hand, the formation of the CO, CO$_2$, and OCN$^-$ species 
for doses above $\approx$ 10 eV/molecule is less attenuated for UV irradiation.
 
In addition to UV photons, 0.8 MeV proton irradiation of 
H$_2$O:CO:CH$_3$OH:NH$_3$ ice also leads to the formation of hexamethylenetetramine (HMT, [(CH$_2$)$_6$N$_4$])
(\cite{Cottin2001}). 
After irradiation of an H$_2$O:NH$_3$:CO = 1:0.6:0.4 ice by 46 MeV Ni ions, an infrared band observed around 1370 cm$^{-1}$ was tentatively assigned to HMT (\cite{Pilling2010a}). But that identification is not supported by other HMT bands; in particular the 
about four times more intense HMT band around 1234 cm$^{-1}$ seems to be absent from their spectrum.    
In the UV and ion irradiation experiments reported here, the infrared bands characteristic of HMT were not detected in the residue spectra. In the UV experiments, only the simultaneous ice deposition and irradiation led to formation of HMT after warm-up of the ice to room temperature, in line with previous works (\cite{MCS2003}), but the experiment where irradiation occurred after the ice deposition did not produce HMT, suggesting that gas phase reactions during irradiation might have played a role in the formation of HMT precursors. Since UV irradiation leads to higher amounts of organic products if H$_2$O is included in the ice mixture (\cite{MCS2003}, and \cite{Oberg2010}), future swift-ion irradiation experiments should incorporate water in the ice to study the possible formation of HMT. In the absence of H$_2$O, the high concentration of formaldehyde, H$_2$CO, formed in the 
NH$_3$:CH$_3$OH = 1:1 photolyzed ice, structures related to polyoxymethylene (POM, [(-CH$_2$O-)$_n$]) are expected; such POM-like structures are formed at the expense of HMT and carboxylic acid salts (\cite{MCS2003}). 

Recent publications reported the identification of HMT precursors formed during warm-up of irradiated ices with a different starting composition. Among other ice mixtures, H$_2$O:CH$_3$OH:NH$_3$ was explored, and the band near 1460 cm$^{-1}$, previously attributed to NH$_4$$^+$, was identified as the protonated ion trimethyletriamine 
(TMTH$^+$, C$_3$H$_{10}$N$_3$$^+$), see Vinogradoff et al. (2013) and references therein. The TMT molecule was identified in residues, along with other N-heterocycles, by gas chromatography coupled to mass spectrometry (\cite{Meierhenrich2005}). If these species were present in our residues, which were formed in the absence of 
H$_2$O,  they did not lead to the formation of HMT in an observable amount. In Table \ref{residue_carriers} we include NH$_4$$^+$ and TMTH$^+$ as the possible carriers of the band near 1454 cm$^{-1}$, although the integrated IR cross-section of TMTH$^+$ would be required for a more quantitative assignment.
 
The reaction network induced by ion- or photo-processing of the CH$_3$OH:NH$_3$ ice mixture, followed by thermal 
processing, can be highly complex. Even the irradiation of pure CH$_3$OH ice with UV photons leads to a complex 
chemistry (\cite{Oberg2009}). It is important to note that the warm-up process followed the same protocol in the UV and the ion experiments reported here because the reactions leading to the more complex species generally occur during heating of the irradiated ice, which allows radicals and ions to recombine and form new molecules. Several selected reactions are provided below, which are expected to play a role in the formation of the detected products. A similar network was proposed earlier to explain the synthesis of UV-photoproducts in ices, whose functional groups are identified by IR spectroscopy, and some individual molecules were detected by chromatography coupled to mass spectrometry  
(\cite{Agarwal1985}, and references therein; the formation of ammonium salts of carboxylic acids was proposed by \cite{MCS2003}). 

Methanol in the ice can lead to the following reactions:

\noindent{CH$_3$OH + ion/UV  $\rightarrow$ CH$_3$$^{\cdot}$ + OH$^{\cdot}$}\\
CH$_3$OH + ion/UV  $\rightarrow$ CH$_2$OH$^{\cdot}$/OCH$_3$ + H$^{\cdot}$ $\rightarrow$ H$_2$CO $\rightarrow$  HCO $\rightarrow$ CO. The so-formed CO can be excited by irradiation and react:\\
CO + ion/UV  $\rightarrow$ C + O / CO*, and CO* + CO   $\rightarrow$  CO$_2$ + C\\
CH$_3$OH + H$^{\cdot}$ $\rightarrow$ CH$_2$OH$^{\cdot}$ + H$_2$, or H$_2$CO + H$^{\cdot}$   $\rightarrow$ CH$_2$OH$^{\cdot}$\\
CH$_3$$^{\cdot}$ + H$^{\cdot}$ $\rightarrow$ CH$_4$\\
HO$^{\cdot}$ + CO   $\rightarrow$ HOOC$^{\cdot}$ $\leftrightarrow$ CO$_2$ + H$^{\cdot}$\\
H$^{\cdot}$ + CO      $\rightarrow$ HCO$^{\cdot}$\\
CH$_2$OH$^{\cdot}$ + HCO$^{\cdot}$ $\rightarrow$ HOCH$_2$CHO (hydroxy-acetaldehyde, formed during warm-up)\\
2 CH$_2$OH$^{\cdot}$ $\rightarrow$ HOCH$_2$CH$_2$OH (ethylene glycol, formed during warm-up, Chen et al. 2013)\\
CH$_2$OH$^{\cdot}$ + COOH$^{\cdot}$   $\rightarrow$ HOCH$_2$COOH (glycolic acid, a carboxylic acid, formed during 
warm-up).

The presence of ammonia in the CH$_3$OH:NH$_3$ ice mixture leads to
 
\noindent{NH$_3$ + ion/UV  $\rightarrow$ NH$_2$$^{\cdot}$ + H$^{\cdot}$}.

The NH$_x$ radicals allow the formation of a variety of new species; most of them are produced during the warm-up of the irradiated ice:

HNCO (from NH$^{\cdot}$ + CO) + NH$_3$ $\rightarrow$ NH$_4^+$OCN$^-$ (during warm-up, \cite{Mispelaer2012} and references therein),

\noindent{NH$_2$$^{\cdot}$ + CO $\rightarrow$ NH$_2$CO$^{\cdot}$}\\
NH$_2$CO$^{\cdot}$ + H$^{\cdot}$ $\rightarrow$ HCONH$_2$ (formamide)\\
2 NH$_2$CO$^{\cdot}$    $\rightarrow$ H$_2$NCOCONH$_2$ (ethanediamide)\\
NH$_2$CO$^{\cdot}$ + NH$_2$$^{\cdot}$  $\rightarrow$ NH$_2$CONH$_2$ (urea, an amide)\\
NH$_2$CO + CO$_2$H$^{\cdot}$ $\rightarrow$ NH$_2$COCO$_2$H (oxamic acid)\\
CH$_2$OH$^{\cdot}$ + CONH$_2$$^{\cdot}$ $\rightarrow$ HOCH$_2$CONH$_2$ (hydroxyacetamide)\\
HOCH$_2$COOH + NH$_3$ $\rightarrow$ [NH$_4^+$][HOCH$_2$COO$^-$] (ammonium salts of carboxylic acids).\\

The release of molecular ice components into the gas phase by swift-ion 
sputtering was an important effect, as shown by the strong increase of the 
gas pressure in the vacuum chamber when the ion beam was on. This desorption is due to electronic sputtering,
which strongly depends on the stopping power (\cite{Seperuelo2009}). This effect was 
not quantified yet for the ice mixtures under investigation here, but it is clearly more efficient than UV-photodesorption;
in the latter case such a high increase in the pressure was not observed. The reason for this is that the
main effect of vacuum-UV radiation of CH$_3$OH:NH$_3$ ice is photodissociation
that induces the formation of new species.    

Interstellar ices are simultaneously subjected to UV photons and ions. 
Several publications have provided an estimate of the photon and ion fluxes that dust grains and icy grain mantles experience in interstellar and circumstellar environments, and the energy deposited on the dust by both types of radiation was  
estimated as well (\cite{Cecchi1992}; \cite{Shen2004}); the recent publication of Islam et al. (2014) provided an update of these results. 
Secondary UV photons are expected to deposit a higher energy dose (typically a few eV/molecule in about one million years) in the ice mantles of external UV-shielded regions (dense clouds) than cosmic ray ions, by a still debated factor of about up to an order of magnitude (see, e.g., Table 3 of \cite{Shen2004}), but cosmic rays can penetrate deeper into ice mantles (most of them pass through interstellar grains).
However, little attention was payed to swift heavy ions since most experiments have dealt with proton or electron bombardment of astrophysical ice analogs. Nevertheless, as mentioned in Sect.~\ref{intro}, these ions are expected to be an important component of cosmic rays in space; in the past five years several experiments simulated their effects in various molecular ice components (\cite{Seperuelo2009}, 2010; \cite{Pilling2010a}, 2010b; \cite{Dartois2013}). We conclude that in our selected ice mixture where both components, methanol and ammonia, are dissociated easily by vacuum-UV photons, similar photon and ion energy doses led to similar chemical reactions, based on the similarity of the IR spectra of the processed ices. In general, it is therefore not possible to use simple astrophysical molecules (detected in ice mantles or in the gas phase, toward, for instance, hot cores or the higher temperature regions in a protosolar nebula disk, where ice mantles sublimate) as tracers of either UV or ion irradiation. Even the direct dissociation of the most stable molecules such as N$_2$ or CO in the ice, which cannot be achieved by the vacuum-UV photons in our experiments, can be attained by extreme-UV or X-ray photons (e.g., Wu et al. 2003, 
and Ciaravella et al. 2012), and therefore the dissociation of these molecules is not intrinsic to ion processing. In addition, to first order, after simultaneous UV/ion processing, a unified organic refractory residue would be formed that is common in the laboratory to UV and ion processing.
  
\section{Conclusion}

\begin{figure}[htbp]
\begin{center}
\includegraphics[width=\columnwidth]{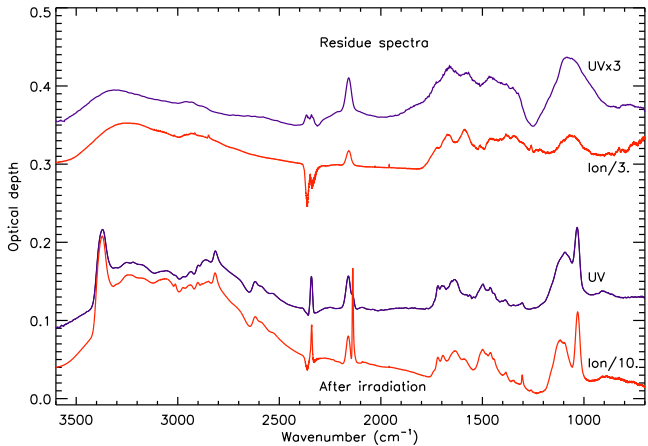}
\caption{Comparison of CH$_3$OH:NH$_3$ = 1:1 ice mixture after UV or 620 MeV ion irradiation. The corresponding residue spectra at room temperature are shown as well.}
\label{fig_comparison}
\end{center}
\end{figure}

Experiments using a similar ice mixture and energy dose (in eV molecule$^{-1}$) 
show that UV photon and swift heavy ion irradiation lead to the formation 
of simple species near 10 K that are common in both experiments for energy doses up to a few tens of eV per molecule, typical of exposures of millions of years in space. At room temperature, both sources of irradiation formed
residues with similar infrared spectral bands, see Table \ref{residue_carriers}. For a better comparison, Fig.~\ref{fig_comparison} presents the spectra of the irradiated CH$_3$OH:NH$_3$ = 1:1 ice mixture using UV or 620 MeV Zn ions, and the corresponding residues that remained at room temperature. The synthesis of these products is expected to follow a complex set of reactions, which were discussed in Sect.~\ref{discussion}. 
This finding has important implications for models of the chemistry in icy grain mantles in space. 
We mentioned in Sect.~\ref{intro} that other works comparing protons and UV photons arrived at a similar conclusion. Both methanol and ammonia are easily dissociated by ions or vacuum-UV photons, forming simple radicals and fragments that react during warm-up. The warm-up rate was the same in our experiments. The immediate question is, therefore, whether a similar chemistry can be expected, as was observed experimentally.  A priori, some differences can be envisioned. Among the products, CO was observed in the irradiated ice of our ion and UV experiments. Similar to N$_2$, see Islam et al. (2014) and references therein for differences between ions and vacuum-UV photons in irradiated ices containing N$_2$, the CO molecule can be directly dissociated by ions but not by UV with photon energies below 10.2 eV, which is the case of the hydrogen UV lamp that we employed. Incoming ions will gradually dissociate the previously formed CO, while a large part of the CO generated by UV will remain intact upon further UV irradiation (not all, because photodesorption of CO operates in the top monolayers (e.g., \cite{MC2010}) and CO can participate in the ice photo- and thermal chemistry by reacting with another UV-excited CO molecule or a different species). Ions can not only dissociate CO ice molecules directly, they can also be sputtered along with other species. Therefore, we did not foresee a similar evolution of the CO abundance curve in the irradiated ice using different sources. As a comparison, clear differences were observed among products formed by either X-ray or UV-irradiation of pure CO ice (\cite{Ciaravella2012}). 


To study the  observed similarity between vacuum-UV and swift ion ice processing, the 
formation of ice irradiation products by swift ions will be explored in the 
near future using different ice mixtures that include H$_2$O. 
If, as is the case for the studied CH$_3$OH:NH$_3$ ice mixture, the 
organic residue formed by swift ion irradiation of ice is similar to that 
obtained by UV photon irradiation, the use of swift ions will allow us to grow 
thicker residues and perform analyses that require higher amounts of material 
than previous studies such as nuclear magnetic resonance (NMR), which will provide an access to 
a complementary identification of functional groups present in residues. 
Furthermore, a chromatographic analysis of thick residues made by swift-ion processing
of ice might also lead to the detection of new molecular components in the organic residues. 

The synthesis of complex organic species results from both UV- and ion-processing experiments followed by warm-up. These species might be present in circumstellar regions and some comets, and were likely delivered to the primitive Earth.     

\begin{acknowledgements}
We are grateful to T. Been, C. Grygiel, T. Madi, I. Monnet and J. M. Ramillon for their invaluable support. This work has been supported by the European Community as an Integrating Activity 'Support of Public and Industrial Research Using Ion Beam Technology (SPIRIT)' under EC contract no. 227012, the Spanish MICINN/MINECO
under projects AYA2011-29375 and CONSOLIDER grant CSD2009-00038.
\end{acknowledgements}

\end{document}